\documentclass[a4paper, 11pt]{article}
\usepackage{amsmath, amsthm,amssymb,mathtools}
\usepackage{times}
\usepackage{enumitem}
\usepackage{graphicx}
\usepackage{xcolor}
\usepackage{subcaption}

\usepackage{mathrsfs}

\renewcommand{\Box}{\rule{1.5mm}{3mm}}

\newtheorem{thm}{Theorem}
\newtheorem{lem}{Lemma}
\newtheorem{prop}{Proposition}
\newtheorem{cor}{Corollary}

\newtheorem{rmk}{Remark}
\newtheorem{defin}{Definition}
\newtheorem{ex}{Example}
\newtheorem{pb}{Problem}

\newcommand{\bp}{\begin{pb}\rm}
\newcommand{\ep}{\end{pb}}
\newcommand{\br}{\begin{rmk}\rm}
\newcommand{\er}{\end{rmk}}
\newcommand{\bdefin}{\begin{defin}\rm}
\newcommand{\edefin}{\end{defin} }
\newcommand{\bex}{\begin{ex}\rm}
\newcommand{\eex}{\end{ex}}

\newcommand{\bthm}{\begin{thm}}
\newcommand{\ethm}{\end{thm}}
\newcommand{\blem}{\begin{lem}}
\newcommand{\elem}{\end{lem}}
\newcommand{\bprop}{\begin{prop}}
\newcommand{\eprop}{\end{prop}}
\newcommand{\bcor}{\begin{cor}}
\newcommand{\ecor}{\end{cor}}

\usepackage{algorithm}
\usepackage{algorithmic}

\usepackage{xargs}                      

\renewcommand{\Box}{\rule{1.5mm}{3mm}}
\setlist[itemize]{noitemsep}

\begin{document}
\begin{center}
{\large \sc Stick graphs: examples and counter-examples}
\bigskip

Irena Rusu

{\it LS2N, University of Nantes, France\\
\small Irena.Rusu@univ-nantes.fr}
\end{center}
\bigskip\bigskip

\begin{center}
\begin{minipage}[h]{13cm}
\paragraph{Abstract}
 {\small\em 
 Grid intersection graphs are the intersection graphs of vertical and horizontal segments in the plane.  When the
 bottom and respectively left endpoints of the vertical and horizontals segments 
 belong to a line with negative slope, the graph is called a Stick graph. Very few   results exist on Stick graphs: only small classes of Stick graphs 
 have been identified; recognizing Stick graphs is an  open problem; and even building examples of graphs that are not 
 Stick graphs is quite tricky. 
 
 In this paper, we first prove that the complements of circle graphs  and of circular arc graphs are Stick graphs.
 Then, we propose two certificates allowing to decide that a graph is not a Stick graph, and use them to build new 
 examples of non-Stick graphs. It turns out that these examples of non-Stick graphs, as well as all those from literature, 
 have long holes. We thus also investigate the place of chordal grid intersection graphs in the hierarchy of classes built 
 around Stick graphs. 
 
 }
 \end{minipage}
 \end{center}

\section{Introduction}

\begin{figure}[t]
 \centering
 \includegraphics[height=9cm]{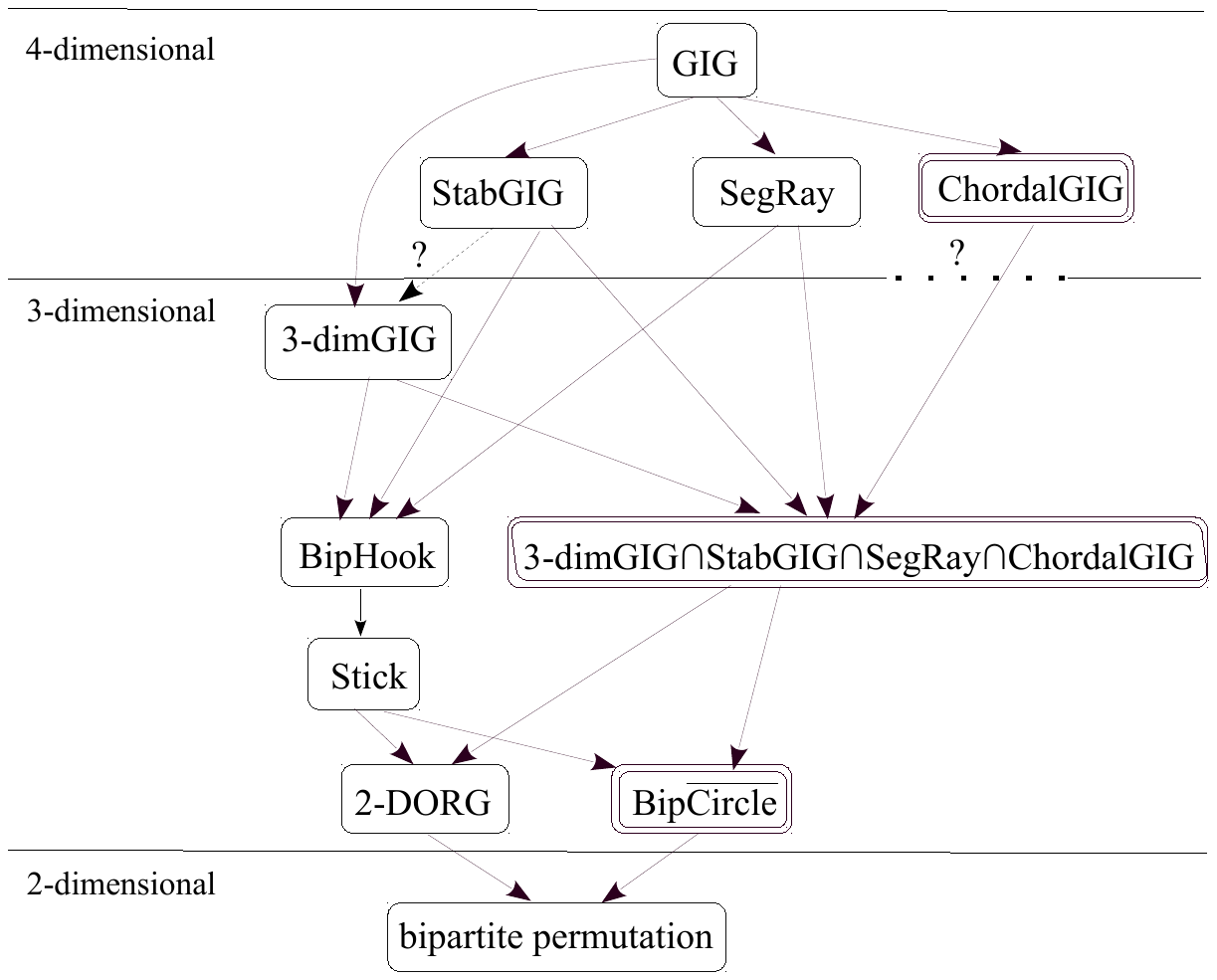}
 \caption{\small Inclusion and overlap relations concerning Stick graphs. Pairs of classes not joined
 by an arrow or by a path are strictly overlapping. 
 Simple rectangular boxes and arrows between them: classes and relations issued from \cite{chaplick2018grid}. 
 Double rectangular boxes and the arrows connected to them: classes and relations studied in this paper. Left
 mark ``?'': open question from \cite{chaplick2018grid}. 
 Right mark ``?'': open question from this paper (see Remark \ref{rem:Chordal3dim} in Section \ref{sect:holes}).} 
 \label{fig:classes}
\end{figure}

Graph classes defined as intersection graphs of geometric objects have been extensively studied. Intersection graphs
of intervals on the real line (called {\em interval graphs}), of arcs on a circle (called {\em circular arc graphs}),
of trapezoids with bases on two fixed parallel lines (called {\em trapezoid graphs}), of chords of a circle
(called {\em circle graphs}), of straight line segments in the plane (called {\em segment graphs}) are only a few examples.
Many of them, together with references and relations between them, may be found in the book \cite{brandstadt1999graph} 
as well as in \cite{deRidder}. Among segment graphs, the class of {\em grid intersection graphs} \cite{hartman1991grid} 
plays a particular role, since it concerns only intersections between a horizontal segment and a vertical segment, 
yielding applications identified  in \cite{chaplick2018grid}. In the same paper, several subclasses of grid intersection graphs are proposed
and their intersections are studied. One of them is the class of Stick graphs, for which the known inclusions are
described in Figure~\ref{fig:classes} (simple rectangular boxes), which is a part of the more general inclusion diagram provided in \cite{chaplick2018grid}.
These classes are defined in Section~\ref{sect:Intro}.

A {\em Stick graph} is the intersection graph of a set $A$ of vertical segments
and a set $B$ of horizontal segments in the plane (respectively called {\em $A$-} and {\em $B$-segments}),
whose bottom and respectively left endpoints lie on a ``ground'' line with slope -1. Each of these
endpoints is called the {\em origin} (which is then an $A$-origin or $B$-origin) of the corresponding segment. 
It is assumed that all the origins are distinct. Otherwise, one may slightly move some of them 
and appropriately modify some segment lengths, without affecting the segment intersections.

Stick graphs, as all the grid intersection graphs, are bipartite graphs. Given a Stick graph,
a representation of it with vertical and horizontal segments as mentioned above is called a {\em Stick representation} of the graph.
The problem of recognizing Stick graphs - denoted {\sc STICK} - requires, given a bipartite graph $G$, to decide 
whether $G$ has a Stick representation or not. {\sc STICK} is an open problem. Variants of {\sc STICK}, assuming
that the order of the $A$-origins and/or the order of the $B$-origins are known, have been considered
and solved in \cite{luca2018recognition, chaplick2019recognizing, rusu2021forced}, including in the case where
the lengths of the segments are known. However, quite little is known about Stick graphs in general, in particular no
structural property is available.
For this reason, the proofs of the relations identified in Figure~\ref{fig:classes} (simple rectangular boxes) take advantage of the 
geometric representation  of each class as a particular grid intersection graph, and not of the structural properties of the 
involved classes,  since for many of them (especially those in the top part of the diagram) such properties are missing.
Whereas the recognition problem is NP-complete for the class GIG \cite{kratochvil1994special}, its hardness is not known
for all the classes in the diagram between GIG (non included) and Stick (included) (see \cite{chaplick2018grid}). 
Note however that the recognition problem for grounded segment graphs, defined similarly with Stick graphs but in which 
the slopes of the segments are arbitrary, is $\exists\mathbb{R}$-complete \cite{cardinal2018}.

Unlike the previous approaches \cite{luca2018recognition, chaplick2019recognizing, rusu2021forced} seeking to test whether a graph is a Stick graph, our
aim is to investigate Stick graphs - and therefore also non-Stick graphs - from the viewpoint of their structure.
More precisely, we look for sufficient structural conditions ensuring that a given graph is a 
Stick graph or, on the contrary, that the graph is not a Stick graph. As a
consequence, we identify well-known classes of graphs that 
are subclasses of Stick graphs and also forbidden subgraphs of Stick graphs.

\paragraph{Outline and contributions.} The paper is structured as follows. We give in Section \ref{sect:Intro} the main definitions.
In Section \ref{sect:classes} we focus on  bipartite complements of circle graphs
and bipartite complements of circular arc graphs, and show that both classes are included in the class of Stick graphs.  
For both of them we propose direct proofs, based on the representation of complements of circle graphs 
(respectively on the representation of circular arc graphs)  
as intersection graphs. For the former class the result is new. For the latter class, an indirect proof of the 
inclusion in the class of Stick graphs exists \cite{chaplick2018grid}, that uses the equivalence between 
bipartite complements of circular arc graphs and the graphs in 2-DORG \cite{shrestha2010orthogonal}. 
Next, in Section~\ref{sect:main}, we identify particular configurations of graphs that 
act as obstructions to the existence of a Stick representation.  Using them, we build several families of graphs that 
are not Stick graphs, and are thus forbidden  subgraphs of Stick graphs. All these examples, as well as the other 
examples presented in \cite{chaplick2018grid} 
and \cite{luca2018recognition}, have a common feature: they contain a hole, {\em i.e.} an induced 
cycle with at least five vertices (however, only  holes with an even number of vertices, thus at least six, 
will appear in bipartite graphs). Therefore, in Section \ref{sect:holes} we study the relationships between the 
chordal grid intersection graphs, which are by definition the grid intersection graphs with no long holes, and 
their relationships with the classes of 
graphs including or included in the class of Stick graphs.  In the 
diagram in Figure~\ref{fig:classes}, we use double rectangular boxes to indicate the classes we studied and
the relations (inclusions and overlaps) we detected with the existing classes.
Section \ref{sect:conclusion} is the conclusion.

\section{Definitions}\label{sect:Intro}

All the graphs we use are undirected and simple, and the notation is classical. A {\em line} is by definition a straight line. 

We define the graph classes in Figure \ref{fig:classes} following \cite{chaplick2018grid}.
{\em Grid intersection graphs} (GIG) are the intersection graphs of vertical and horizontal segments in the plane,
where two vertical (respectively two horizontal) segments never intersect. When such a representation (called a {\em GIG-representation})
exists for a graph $G$, such that - moreover - all the segments of the representation are intersected by the same line, the
graph is called a {\em stabbable grid intersection graph} (StabGIG).  A graph admitting a GIG-like representation
where the vertical segments are replaced by vertical {\em rays} (i.e. half-lines) oriented in the up direction is
called a {\em segment-ray graph} (SegRay). If, moreover, in the GIG-like representation the horizontal segments are also
replaced by horizontal rays oriented in the right direction, the graph is called a {\em two direction orthogonal ray graph}
(2-DORG). Going a step further, a {\em hook} is a couple of segments, a vertical one and a horizontal one, that share the
same bottom and respectively left endpoint. This point is the {\em center} of the hook. {\em Bipartite hook graphs}
(BipHook) are the intersection graphs of hooks whose centers lie on the same 
line with slope -1 and such that the resulting graph is bipartite. Finally, {\em bipartite permutation graphs} are defined  
as the intersection graphs of  segments with one endpoint on each of two given parallel lines such that the resulting graph is bipartite.

The notion of {\em dimension} of a bipartite graph is needed to complete the background necessary to understand the diagram in 
Figure \ref{fig:classes}. Given a bipartite graph $G=(A\cup B,E)$, let the partial order $\leq_G$ on $A\cup B$ be defined 
such that $A$ ($B$) is the set of minimal (maximal) elements in $A\cup B$ and, for each
$a\in A$ and $b\in B$, the relation $a\leq_G b$ holds if and only if $ab\in E$. The {\em dimension} of $G$ - which is 
unchanged if classes $A$ and $B$ are switched - is then
the minimum $k$ such that a set $\{\leq_1, \leq_2, \ldots, \leq_k\}$ of linear extensions of $\leq_G$ exists with the property
that $a\leq_G b$ iff $a\leq_i b$ for every $i$ with $1\leq i\leq k$. Such a set is called a {\em realizer} of the partially
ordered set $(A\cup B, \leq_G)$. 
In the diagram, the $i$-dimensional regions contain the classes whose graphs have dimension at most $i$, and contain at least one graph
of dimension equal to $i$.
The class $3$-dim GIG thus contains the grid intersection graphs of dimension $3$ or less.

We focus now on Stick graphs. The bipartite graph for which a Stick representation is searched is always denoted by $G=(A\cup B,E)$,
with $|A|=n$ and $|B|=m$. The same notation $a_i$, $1\leq i\leq n$, is used for a given vertex in $A$ and for the origin of 
the vertical segment representing it in any Stick representation of $G$. Thus each $A$-segment is identified by its origin $a_i$ 
and by its {\em tip} $T(a_i)$, {\em i.e.} the top endpoint of the $A$-segment with origin $a_i$.  
Similar definitions and notation hold for 
the vertices in $B$. As initially done for the origins only, we may consider that all the origins and the tips are distinct.

\begin{figure*}[t!]
    \hspace*{-0.9cm}
    \scalebox{1.15}{\begin{subfigure}[t]{0.62\textwidth}
        \centering
        \vspace*{-1.5cm}
        \includegraphics[angle=270, width=0.9\textwidth]{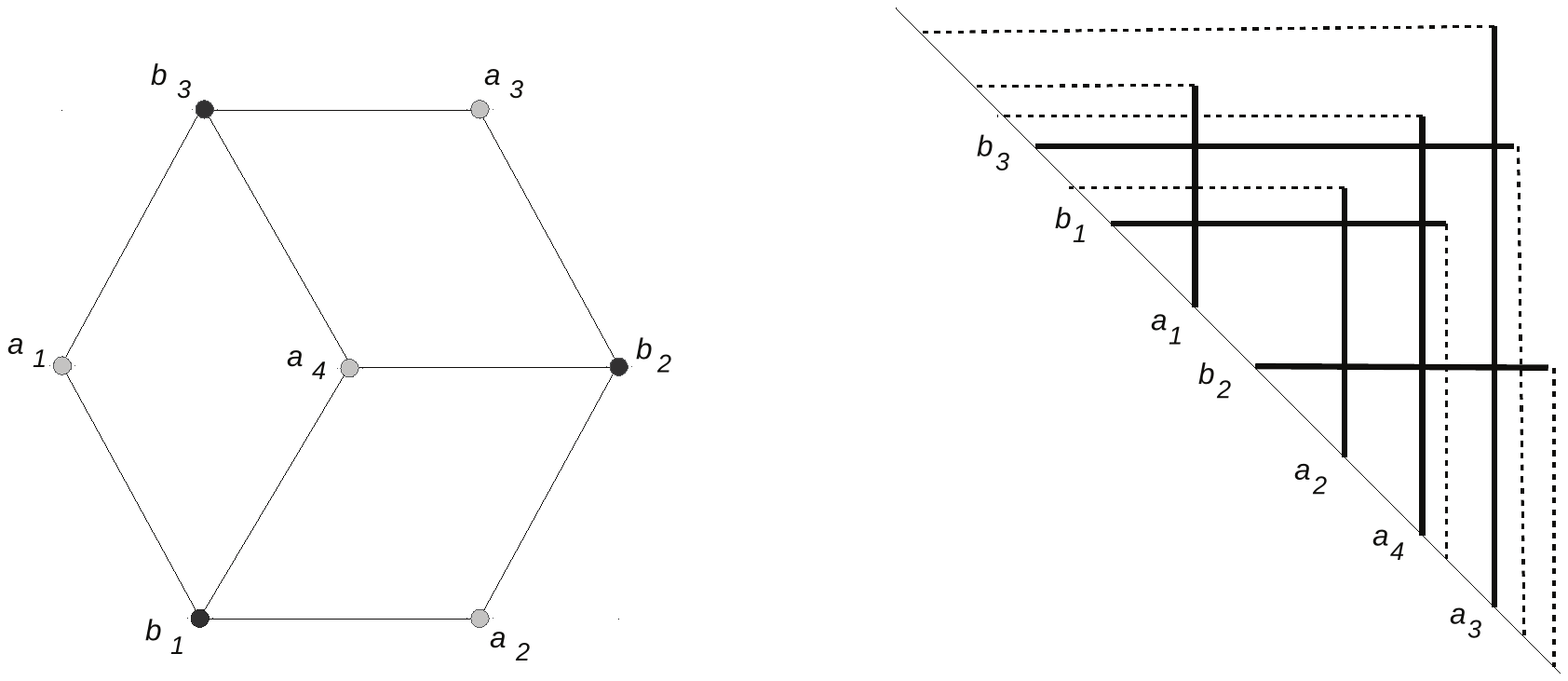}
        \caption{}
    \end{subfigure}
    \begin{subfigure}[t]{0.33\textwidth}
        \centering
        \vspace*{-0.5cm}
        \includegraphics[width=0.95\textwidth]{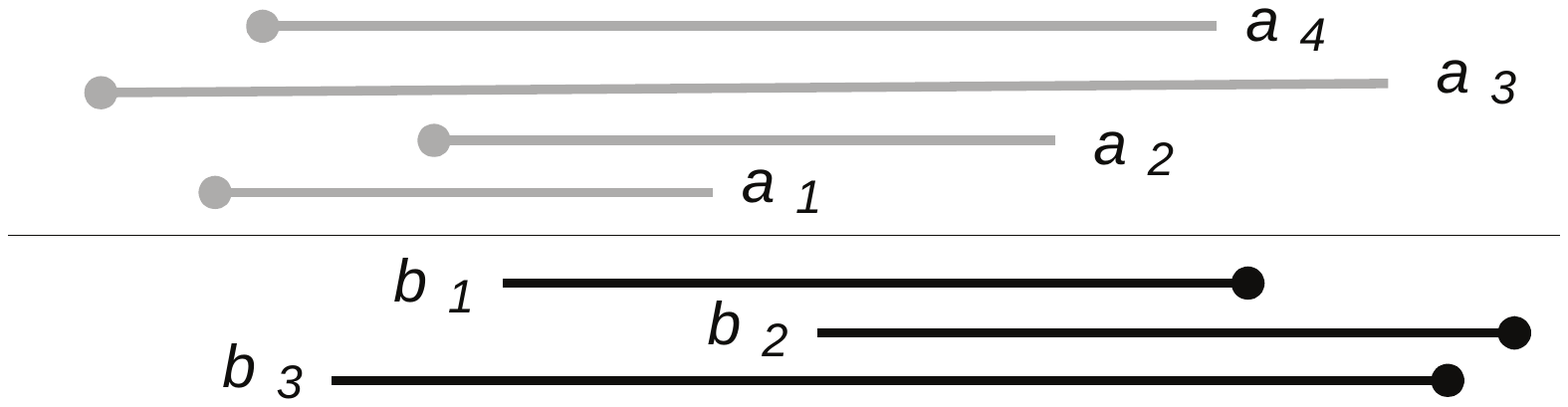}
        \vspace*{2cm}
        \caption{}   
    \end{subfigure}}
    \caption{\small  a) A Stick graph and its Stick representation, with continuous lines. The dotted lines indicate how to compute the tips.
    b) The resulting flat Stick representation, where the plain circles represent the tips. }
    \label{fig:examplesStickflat}
    \end{figure*}

For any two points $x,y$ of the ground line,  we write $x \prec y$ if $x$ is above $y$. A Stick representation of a graph $G$
is thus identified by a linear ordering $\prec$ of the set of origins and tips corresponding to the vertices in $G$.
The following observation is easy (and similar
to those made for other classes, {\em e.g.} for max-point tolerance graphs \cite{catanzaro2017max}):
\bigskip

\noindent{\bf Observation.} Segments $A_i$ and $B_j$ intersect in a Stick representation  
iff  $T(a_i)\prec b_j\prec a_i\prec T(b_j)$. We then say that they {\em overlap}.
\bigskip

\noindent{\bf Conventions.} In Sections \ref{sect:classes} and \ref{sect:main}, we use this {\em flat} Stick representation of Stick 
graphs,  where the ground line is assumed to be the real line (with ``top'' becoming ``left'', and ``bottom'' becoming ``right'')
and each vertical (horizontal) segment is represented as an interval $[T(a_i),a_i]$ ($[b_j, T(b_j)]$) such that 
a vertical and a horizontal segment overlap if and only if they define an edge in the Stick graph.

\bex
Figure \ref{fig:examplesStickflat} shows a Stick graph $G$ together with its standard and flat Stick representations.
\eex

\section{Classes of Stick graphs}\label{sect:classes}

In this section we show that bipartite complements of circle graphs (class ${\rm Bip}\overline{\rm Circle}$) are Stick graphs,
and give an alternative proof that bipartite complements of circular arc graphs (equivalent with 2-direction orthogonal ray graphs, 
or 2-DORG) are Stick graphs.

\bthm
${\rm Bip}\overline{\rm Circle} \subsetneq {\rm Stick}$.
\label{thm:complementscircle}
\ethm

{\bf Proof.} In \cite{esperet2020bipartite}, the authors give an alternative proof of a result by Bouchet
\cite{bouchet1999bipartite}. This result states that every graph that is a bipartite 
complement of a circle graph is also a circle graph. Given a graph $G=(A\cup B, E)$ which
is a bipartite complement of a circle graph, the proof of Theorem 1 in \cite{esperet2020bipartite} shows that
$G$ is the intersection graph of a set  of chords with pairwise distinct endpoints in a circle $C$
and such that all the chords intersect a line $l$. The line $l$ may be
assumed horizontal such that $p$ is the leftmost point of the circle and $q$ its rightmost point. 
Assume $G$ is connected (otherwise make the same reasoning for each connected component)
and let $S_A$, $S_B$ be the sets of chords corresponding to the vertices in $A$ and $B$.
Then each chord $[xy]$ in $S_A$ ($[zt]$ in $S_B$) has its endpoint $x$ (respectively $z$)
between $p$ and $q$ in the top half of the circle $C$, and its endpoint $y$ (respectively $t$)
in the bottom half of the circle $C$. Without loss of generality we assume that the endpoint $y$ or $t$
closest to $p$ is the endpoint $y_0$ of a chord $[x_0y_0]$.

We show that:
\medskip

\noindent $(P)$ Each pair of intersecting chords $[xy]\in S_A$ and $[zt]\in S_B$ has the
property that $p, z, x, q, t,y,p$ are in this order along $C$ in the clockwise direction.
\medskip

Note that $[x_0y_0]$ and each chord intersecting it satisfy this property. If, by contradiction,
the property is false, then there exists a pair of intersecting chords $[x'y']\in S_A$ and $[z't']\in S_B$ such
that $p, x', z', q, y',t',p$ are in this order along $C$ in the clockwise direction. 
Since $G$ is connected, there exists:
\begin{itemize}
\item either a sequence $[x_0y_0], [z_0t_0], [x_1y_1], [z_1t_1], \ldots,  [x_uy_u], [z_ut_u]$
of chords with $[x_uy_u]=[x'y']$ and $[z_ut_u]=[z't']$
\item or a sequence $[x_0y_0], [z_0t_0], [x_1y_1], [z_1t_1], \ldots,  [x_uy_u], [z_ut_u],[x_{u+1}y_{u+1}] $
of chords with $[x_{u+1}y_{u+1}]=[x'y']$ and $[z_ut_u]=[z't']$.
\end{itemize}
In both cases, we may assume that the sequence is as short as possible between $[x_0,y_0]$ and 
the pair of chords $[x'y']$ and $[z't']$. Now, in the former
case we have that $p, z_{u-1}, x_u, q, t_{u-1},y_u,p$ (on the one hand) and $p, x_u, z_u, q, y_u,t_u,p$ (on the other hand)
are in this order along $C$, which implies that $p, z_{u-1}, x_u, z_u, q, t_{u-1},$ $y_u,t_u, p$ are in this
order along $C$ and thus $[z_u,t_u]$ and $[z_{u-1},t_{u-1}]$ intersect, a contradiction. In the
latter case, we similarly have that $p, z_u, x_u, q, t_u,y_u,p$ (on the one hand) and $p, x_{u+1}, z_u, q, y_{u+1},t_u,p$ 
(on the other hand) are in this order along $C$, which implies that $p, x_{u+1}, z_u, x_u, q, y_{u+1}, t_u,y_u,p$
are in this order along $C$ and thus $[x_u,y_u]$ and $[x_{u+1},y_{u+1}]$ intersect, a contradiction. Thus Property (P)
holds.

Now, let $q^{l}$ and $q^r$ be two points of $C$ situated immediately before and immediately after $q$ in
clockwise direction, so that no endpoint of a chord exists between $q^l$ and $q^r$. Cut the circle at 
the point $q$ and unfold it such that it becomes a straight horizontal line with left endpoint $q^l$ and
right endpoint $q^r$. Each chord in $S_A$ (respectively $S_B$) becomes an interval in $A$ (respectively $B$)
on this line (which represents a segment of the real line). Then, using Property $(P)$,
each pair of intersecting intervals $[x,y]\in A$ and $[z,t]\in B$ has the
property that $q^l, x, z, p, y, t, q^r$ are in this order along the real line. Indeed, the intersecting chords
became intersecting intervals, and non-intersecting chords became intervals included in each other.
Therefore the intervals in $A$ and $B$ provide a flat Stick representation for $G$. 

The inclusion of the class ${\rm Bip}\overline{\rm Circle}$ into Stick is strict, as shown by $C_6$. $\Box$

\br
Note that bipartite permutation graphs are included in  ${\rm Bip}\overline{\rm Circle}$. This affirmation results from
the following two facts: a graph is a permutation graph if and only if its complement is a permutation graph 
\cite{pnueli1971transitive}, and permutation graphs are circle graphs \cite{brandstadt1999graph}.
\er

In \cite{shrestha2010orthogonal}, the study of 2-DORG reveals  - using forbidden submatrices of the adjacency matrix -  
that the graphs in the class 2-DORG are exactly the bipartite complements of circular arc graphs. As 
2-DORG $\subsetneq {\rm Stick}$ \cite{chaplick2018grid}, the result below follows. 
We propose here an alternative proof, using the circular representation of $\overline{G}$ and yielding a Stick representation
of $G$.

\begin{thm}
 The bipartite complements of circular arc graphs are Stick graphs. 
\label{thm:bicompcircarc}
 \end{thm}

{\bf Proof.} Let $G=(A\cup B, E)$ be a bipartite complement of a circular arc graph. Then its complement $\overline{G}$ is
a circular arc graph whose vertices are covered by two disjoint cliques, induced by $A$ and $B$. As remarked in \cite{feder2003bi} 
based on a result 
from \cite{spinrad1988circular}, since $\overline{G}$ is  a circular arc graph with clique cover number two  then
$\overline{G}$ has a representation as follows. 
The circle $C$ has top point $p$, rightmost point $r$, bottom point $q$ and leftmost
point $s$. The family of arcs of the circle  representing $\overline{G}$ is 
$\mathcal{K}=S\cup T$, where $S=\{S_v\, |\, v\in A\}$ and $T= \{T_w\, |\, w\in B\}$, such that: each arc $S_v$ contains
$s$ and $q$ but neither $p$ nor $r$; each arc $T_w$ contains $p$ and $r$ but neither $s$ nor $q$;
arcs $S_v$ and $T_w$ intersect iff $vw$ is an edge of $\overline{G}$. The endpoints of the arcs
may be considered as distinct.

Let $C^1$ and $C^2$ be respectively the top half of the circle (containing $p$) and 
the bottom half of the circle (containing $q$). For each arc $S_v$, denoted by $(xy)$ in order to indicate its endpoints, 
such that $x\in C^1$ and $y\in C^2$, let $y'$ be the point of the circle that is situated symmetrically to $y$
with respect to the horizontal line $sr$. For each arc $T_w$, denoted by $(zt)$,   such that $z\in C^1$
and $t\in C^2$, let $t'$  be the point of the circle that is situated symmetrically to $t$
with respect to the horizontal line $sr$. Then $x,z,y',t'\in C^1$. We show that the
sets of intervals $A=\{[x,y']\, |\, (xy)\in S, x\in C^1, y\in C^2\}$ and 
$B=\{[z,t']\, |\, (zt)\in T, z\in C^1, t\in C^2\}$ define a flat Stick representation of $G$ (the
half-circle $C^1$ is then assumed to represent the horizontal line). 

Let $(xy)\in S$ and $(zt)\in T$ be two arcs representing the vertices $v$ and $w$ of $G$.

{\bf Case 1.}  $(xy)$ and $(zt)$ do not intersect. Then $s, x, z,p, r, t, y, q$ appear in this 
order in the clockwise direction along $C$, so that in $C^1$ the same direction but using now $y'$ and $t'$ 
gives the order $s, x, z, p, y', t',r$. The intervals $[x,y']$ and $[z,t']$ overlap with $[x,y']$ to the left of $[z,t']$.

{\bf Case 2.}  $(xy)$ and $(zt)$ intersect only between $s$ and $p$. Then  $s, z, x,p, r, t, y, q$ appear in this 
order in the clockwise direction along $C$, and we deduce that  $s, z, x, p, y', t',r$ appear in this order
in $C^1$ in the clockwise direction. The interval $[x,y']$ is contained in the interval $[z,t']$.

{\bf Case 3.}  $(xy)$ and $(zt)$ intersect only between $q$ and $r$. Then $s, x, z,p, r, y, t, q$ appear in this 
order in the clockwise direction along $C$, so that the order using $y'$ and $t'$ is now  $s, x, z, p, t', y',r$. 
The interval $[z,t']$ is contained in the interval $[x,y']$.

{\bf Case 4.}  $(xy)$ and $(zt)$ intersect both between $s$ and $p$, and between $q$ and $r$. Then $s, z, x,p, r, y, t, q$ appear in this 
order in the clockwise direction along $C$. We deduce that the order using $y'$ and $t'$ is   $s, z, x, p, t', y',r$. 
The interval $[z,t']$ intersects the interval $[x,y']$, with $[z,t']$ to the left of $[x,y']$. 

In conclusion, the order $x, z, y',t'$ occurs iff $(xy)$ and $(zt)$ do not intersect (in case 1), which happens
iff in $\overline{G}$ the vertices $v$ and $w$ corresponding respectively to $(xy)$ and $(zt)$ are not adjacent.
But that means $[x,y']$ and $[z,t']$ overlap (in this order from left to right) iff $v,w$ are adjacent in $G$,
thus we have a flat Stick representation of $G$. $\Box$

 

\section{Obstructions and new examples of non-Stick graphs}\label{sect:main}

\begin{figure*}[t!]
    \vspace*{-2cm}
    \begin{subfigure}[t]{0.30\textwidth}
        \includegraphics[width=0.9\textwidth]{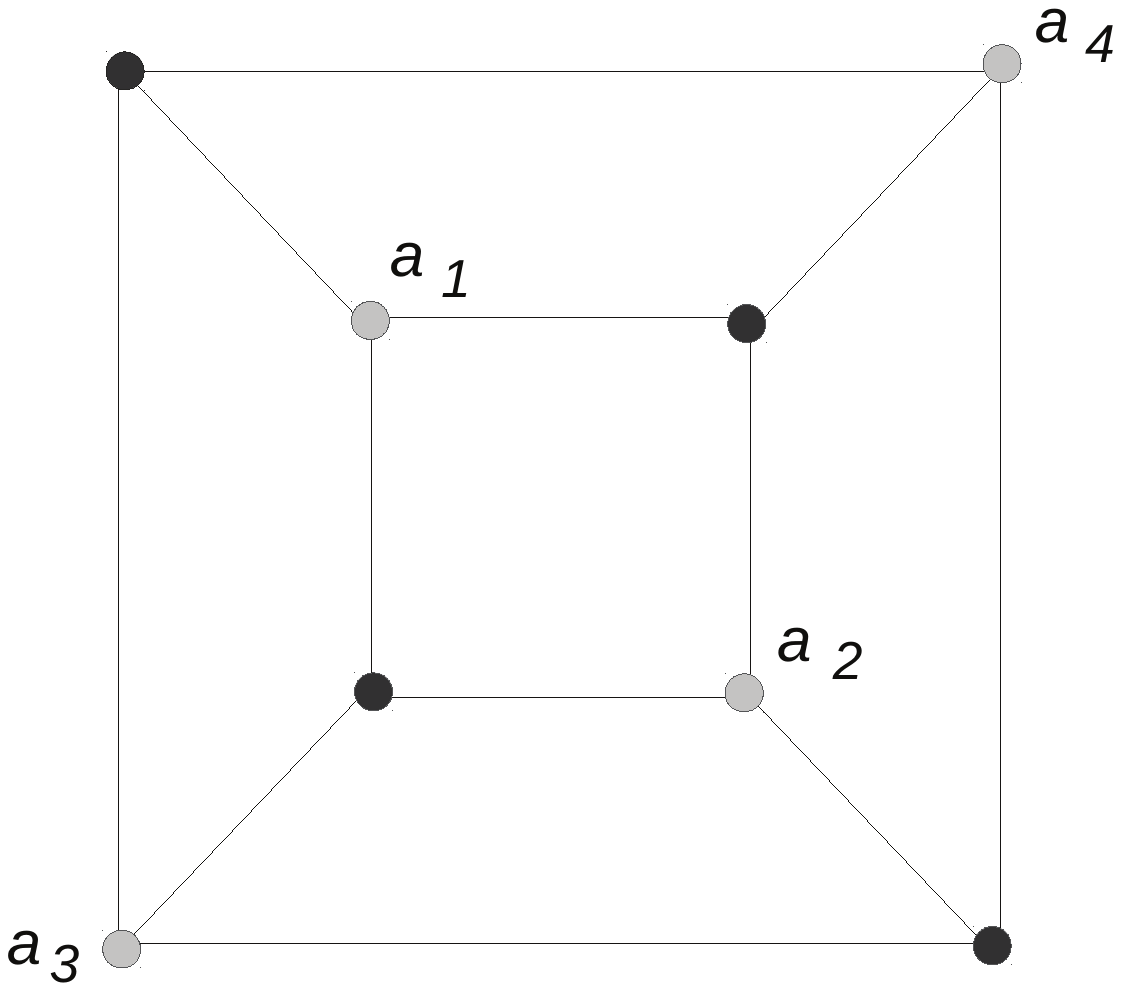}
        \vspace*{-0.5cm}
        \caption{}
    \end{subfigure}
    \hspace*{-2cm}\begin{subfigure}[t]{0.43\textwidth}
        \vspace*{-5.2cm}
        \hspace*{1cm}\includegraphics[width=0.95\textwidth]{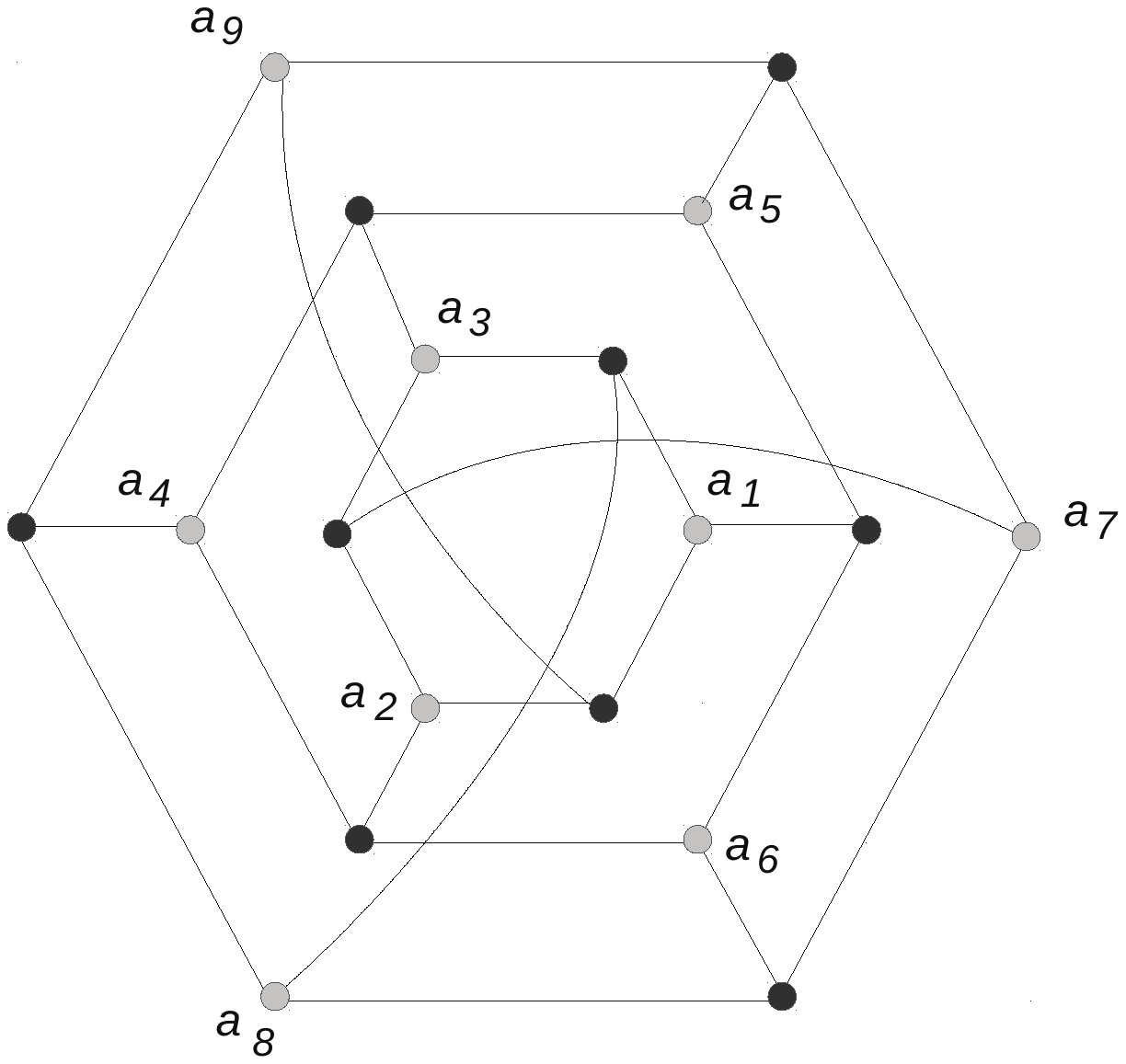}
        \caption{}   
    \end{subfigure}
       \begin{subfigure}[t]{0.4\textwidth}
         \centering
         \vspace*{-5.5cm}
         \includegraphics[width=0.9\textwidth]{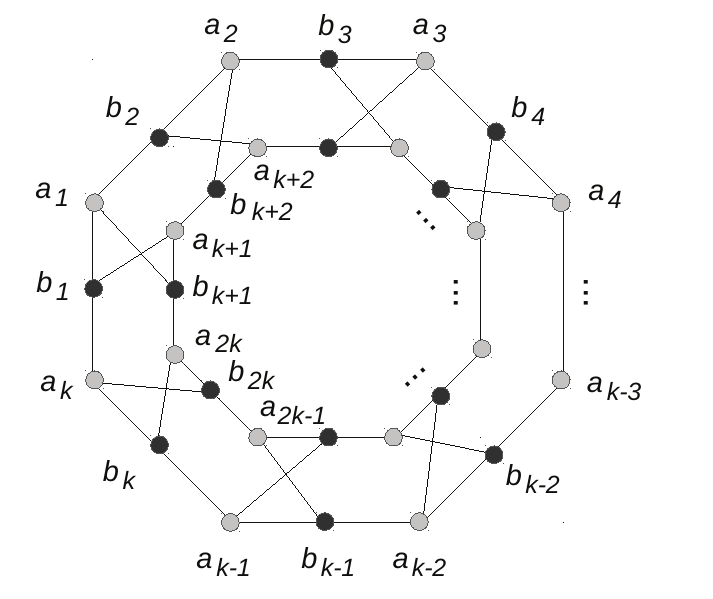}
         \caption{\small Graph $J_k$, for odd $k$ only.}
     \end{subfigure}
    \caption{\small Forbidden subgraphs of Stick Graphs.}
    \label{fig:examples}
    \end{figure*}

 
In Figure \ref{fig:examples}, we give several examples of graphs that are not Stick graphs. The graph in Figure \ref{fig:examples}(a) 
is the smallest graph with this
property, proposed in \cite{luca2018recognition}. The two other graphs are new.

Showing that a graph is not a Stick graph usually needs  a case-by-case proof eliminating every possibility to obtain a Stick 
representation (or equivalently an appropriate ordering of the adjacency matrix, see \cite{luca2018recognition}).
The results presented in this section may allow us to avoid these fastidious checking for some classes of graphs,
as is the case for the graphs in Figure \ref{fig:examples} (see Examples \ref{ex:exampleLuca}-\ref{ex:Jk}), 
by applying certificates that attest that a graph is not a Stick graph.

When two vertices $a_v,a_t\in A$ are fixed in the bipartite graph $G=(A\cup B,E)$, we
define:

$$\mathcal{B}_1=N(a_t)-N(a_v) \text{ and }  \mathcal{B}_2=N(a_t)\cap N(a_v).$$

Moreover, for each $a_j$ with $j\neq v,t$, we define $N_i(a_j)=N(a_j)\cap \mathcal{B}_i$ ($i=1,2$) and say that $a_j$ is a
$1$-{\em witness} (resp. a $12$-{\em witness})   of $a_v,a_t$ if $N_1(a_j)\neq \emptyset \text{ and } N_2(a_j)=\emptyset$ (respectively
if $N_1(a_j)\neq \emptyset$ $\text{ and } N_2(a_j)\neq \emptyset$). Recall that $T(a_i)$ is the notation for the tip of $a_i$, and a
Stick representation $\prec$ is a total order on the set of origins and tips.

\begin{thm}
Let $G=(A\cup B, E)$ be a Stick graph and let $\prec$ be a Stick representation of it. Let $A=\{a_1, a_2, \ldots, a_n\}$ such that $T(a_i)\prec T(a_j)$ iff $i<j$.
Let $a_v,a_t\in A$ such that $v<t$, and 
assume $\mathcal{B}_2\neq \emptyset$.  Then:

\begin{enumerate}[itemsep=0pt]
 \item[$(i)$] the sets in the collection  $\mathcal{C}_1=\{N_1(a_j)\,|\, 1\leq j<t \text{ and } j \text{ is a 1-witness of } a_v,a_t\}$
 are linearly ordered by inclusion.
 
  \item[$(ii)$] the sets in the collection  $\mathcal{C}_2=\{N_2(a_j)\,|\, 1\leq j<t, \text{ and } j \text{ is a 12-witness of } a_v,a_t\}$
 are linearly ordered by inclusion.
\end{enumerate}
\label{thm:main}
\end{thm}

\begin{proof}
Since $v<t$, we have $T(a_v)\prec T(a_t)$.

 $i)$ Let $j\neq k$ such that $N_1(a_j), N_1(a_k)\in \mathcal{C}_1$. We show by contradiction that $N_1(a_j)\subseteq N_1(a_k)$ or 
 $N_1(a_k)\subseteq N_1(a_j)$. To this end, assume that the contrary holds and consider $b_2\in \mathcal{B}_2$, $b_1\in N_1(a_j)\setminus 
 N_1(a_k)$ and $b'_1\in N_1(a_k)\setminus N_1(a_j)$. 
 \medskip
 
  {\bf Case $a_v\prec a_t$.} We successively deduce:
 
 \begin{itemize}[itemsep=0pt]
 \item[$\bullet$] $b_2\in \mathcal{B}_2$ implies that $ T(a_v)\prec T(a_t)\prec b_2\prec a_v \prec a_t\prec T(b_2)$. \hfill (1)
 
 \item[$\bullet$] $b_1\in \mathcal{B}_1$ implies that $a_v\prec b_1\prec a_t\prec T(b_1)$. \hfill (2)
 
 \item[$\bullet$] $j<t$ and $a_jb_1\in E$ imply that $T(a_j)\prec T(a_t)\prec b_2\prec a_v\prec b_1\prec a_j\prec T(b_1)$. \hfill (3)
 
 \item[$\bullet$] $a_jb_2\not\in E$ implies that $T(b_2)\prec a_j$ and thus $b_2\prec a_v\prec b_1\prec a_t\prec T(b_2)\prec a_j\prec T(b_1)$. \hfill (4)
\end{itemize}

 Moreover, deductions (2')-(4') similar with (2)-(4) are obtained with $(a_k,b'_1)$ instead of $(a_j, b_1)$. Assume without loss of 
 generality that $a_j\prec a_k$.
 Then, using (1)-(4) and (2')-(4') we deduce that $T(a_j)\prec T(a_t)\prec b_2\prec a_v\prec b'_1\prec a_t\prec  T(b_2)\prec a_j\prec a_k\prec T(b'_1)$. This implies
 $a_jb'_1\in E$, thus $b'_1\in N_1(a_j)$, a contradiction.
 \medskip
 
 {\bf Case $a_t\prec a_v$.}  We successively deduce:
 
 \begin{itemize}[itemsep=0pt]
 \item[$\bullet$] $b_2\in \mathcal{B}_2$ implies that $ T(a_v)\prec T(a_t)\prec b_2\prec a_t \prec a_v\prec T(b_2)$. \hfill (5)
 
 \item[$\bullet$] $b_1\in \mathcal{B}_1$ implies that $T(a_v)\prec T(a_t)\prec b_1\prec a_t\prec T(b_1)\prec a_v$. \hfill (6)
 
 \item[$\bullet$] $j<t$ and $a_jb_1\in E$ imply that $T(a_j)\prec T(a_t)\prec b_1\prec a_j\prec T(b_1)$. \hfill (7)
 
 \item[$\bullet$] $a_jb_2\not\in E$ implies that $a_j\prec b_2$ and thus $T(a_t)\prec b_1\prec a_j\prec b_2\prec a_t\prec T(b_1)$. \hfill (8)
\end{itemize}

 Moreover, deductions (6')-(8')in similar with (6)-(8) are obtained with $(a_k,b'_1)$ instead of $(a_j, b_1)$. Assume without loss of generality
 that $a_j\prec a_k$.
 Then, using (5)-(8) and (6')-(8') we deduce that $T(a_k)\prec T(a_t)\prec b_1\prec a_j\prec a_k\prec b_2\prec a_t\prec T(b_1)$. This implies
 $a_kb_1\in E$, thus $b_1\ N_1(a_k)$, a contradiction. 
 \bigskip
 
 $ii)$ The proof for this affirmation is similar to the proof of affirmation $(i)$. Let $j\neq k$ such that 
 $N_2(a_j), N_2(a_k)\in \mathcal{C}_2$. We show by contradiction that $N_2(a_j)\subseteq N_2(a_k)$ or 
 $N_2(a_k)\subseteq N_2(a_j)$. To this end, assume that the contrary holds and consider 
 $b_1\in N_1(a_j)$, $b_2\in N_2(a_j)\setminus N_2(a_k)$,  $b'_1\in N_1(a_k)$ and 
 $b'_2\in N_2(a_k)\setminus N_2(a_j)$. 
 \medskip
 
  {\bf Case $a_v\prec a_t$.} We successively deduce (note that affirmations (9)-(11) are identical to affirmations (1)-(3)):
 
 \begin{itemize}[itemsep=0pt]
 \item[$\bullet$] $b_2\in \mathcal{B}_2$ implies that $ T(a_v)\prec T(a_t)\prec b_2\prec a_v \prec a_t\prec T(b_2)$. \hfill (9)
 
 \item[$\bullet$] $b_1\in \mathcal{B}_1$ implies that $T(a_v)\prec T(a_t)\prec a_v\prec b_1\prec a_t\prec T(b_1)$. \hfill (10)
 
 \item[$\bullet$] $j<t$ and $a_jb_1\in E$ imply that $T(a_j)\prec T(a_t)\prec b_2\prec a_v\prec b_1\prec a_j\prec T(b_1)$. \hfill (11)
 
 \item[$\bullet$] $a_jb_2\in E$ implies that $T(a_j)\prec b_2\prec a_j\prec T(b_2)$. \hfill (12)
\end{itemize}

 Moreover, deductions (9')-(12') similar with (9)-(12) are obtained with $(a_k,b'_1, b'_2)$ instead of $(a_j, b_1,b_2)$.
 Then, using (9)-(12) and (9')-(12') we first deduce that $T(a_k)\prec T(a_t)\prec b_2 \prec a_v\prec b'_1\prec a_k$. Now, 
  $a_k\prec T(b_2)$ implies  $a_kb_2\in E$, a contradiction with the hypothesis$b_2\in N_2(a_j)\setminus N_2(a_k)$.  Then
  we have $T(b_2)\prec a_k$, thus $T(a_j)\prec T(a_t)\prec b'_2\prec a_v\prec b_1\prec a_j\prec T(b_2)\prec a_k\prec T(b'_2)$. 
  But this implies $a_jb'_2\in E$, another contradiction.
 \medskip
 
 {\bf Case $a_t\prec a_v$.}  We successively deduce (note that affirmations (13)-(15) are identical to affirmations (5)-(7)):
 
 \begin{itemize}[itemsep=0pt]
 \item[$\bullet$] $b_2\in \mathcal{B}_2$ implies that $ T(a_v)\prec T(a_t)\prec b_2\prec a_t \prec a_v\prec T(b_2)$. \hfill (13)
 
 \item[$\bullet$] $b_1\in \mathcal{B}_1$ implies that $T(a_v)\prec T(a_t)\prec b_1\prec a_t\prec T(b_1)\prec a_v$. \hfill (14)
 
 \item[$\bullet$] $j<t$ and $a_jb_1\in E$ imply that $T(a_j)\prec T(a_t)\prec b_1\prec a_j\prec T(b_1)$. \hfill (15)
 
 \item[$\bullet$] $a_jb_2\in E$ implies that $T(a_j)\prec T(a_t)\prec b_2\prec a_j\prec T(b_2)$. \hfill (16)
\end{itemize}

 Moreover, deductions (13')-(16') similar with (13)-(16) are obtained with $(a_k,b'_1,b'_2)$ instead of $(a_j, b_1,$ $b_2)$.
 Then, using (13)-(16) and (13')-(16') we deduce that $T(a_j)\prec T(a_t)\prec b_1\prec \prec a_j\prec T(b_1)\prec a_v\prec T(b'_2)$. If
 $b'_2\prec a_j$, then  $a_jb'_2\in E$, a contradiction. Otherwise, $a_j\prec b'_2$ implies
 $T(a_k)\prec T(a_t)\prec b_2\prec a_j\prec b'_2\prec a_k\prec T(b'_1)\prec a_v\prec T(b_2)$. But this means
 $a_kb_2\in E$, another contradiction.
 \end{proof}
 
Theorem \ref{thm:main} allows us to deduce the following result:

\bcor
Let $G=(A\cup B, E)$ be a bipartite graph with $A=\{a_1, a_2, \ldots, a_n\}$. Assume that, for each vertex $a_t\in A$, there exists a vertex
$a_v\in A$, with $v\neq t$,   such that $\mathcal{B}_2\neq\emptyset$ and there exist integers 
$j\neq k$ such that:
\begin{itemize}
 \item[$(i)$] either $a_j,a_k$ are 1-witnesses of $a_v,a_t$ such that $N_1(a_j), N_1(a_k)$ strictly overlap, 
 \item[$(ii)$] or $a_j,a_k$ are 12-witnesses of $a_v,a_t$ such that $N_2(a_j), N_2(a_k)$ strictly overlap.
 
 \end{itemize}
 
%

\noindent Then $G$ is not a Stick graph.
\label{cor:notStick}
\ecor

{\bf Proof.} Assume by contradiction that $G$ is a Stick graph, and renumber its vertices
with $a_1, \ldots, a_n$ in increasing order of their tips. Then $T(a_n)$
is the rightmost tip. By hypothesis, there exists a vertex $a_v$ such that $a_n,a_v$ 
possess witnesses $a_j,a_k$ satisfying one of the properties $(i)$ and  $(ii)$. Then $j,k<n$
and thus Theorem~\ref{thm:main} is contradicted. $\Box$
\bigskip

We illustrate the use of this corollary on the three examples in Figure \ref{fig:examples}.

\bex
The graph in Figure \ref{fig:examples}(a) was proposed in \cite{luca2018recognition} and  has the property that {\it each} pair of vertices
$a_v, a_t$ admits two $12$-witnesses (the two remaining $A$-vertices) whose $N_2$-neighborhoods strictly overlap.
Corollary \ref{cor:notStick} thus applies with case $(ii)$ for all vertices $a_t\in A$.
\label{ex:exampleLuca}
\eex

\bex
The graph in Figure \ref{fig:examples}(b) is not a Stick graph either. 
For each vertex $a_t$ ($t=1,2,3$) and each vertex $a_v$ such that $N(a_t)\cap N(a_v)\neq \emptyset$
there exist two disjoint paths with four edges joining $a_t$ and $a_v$. The two $A$-origins 
in these paths, which are not endpoints, are $1$-witnesses of $a_v,a_t$
whose $N_1$-neighborhoods  are strictly overlapping. Corollary \ref{cor:notStick} thus applies with 
case $(i)$ for all vertices $a_t\in A$. Note however that this graph is not a minimal
non-Stick graph, since the induced subgraph obtained by removing the vertices of the outer cycle
$C_6$ and their incident edges is not a Stick graph either. The proof for this minimal graph 
is done - as usual - by a case study, since our corollary does not apply for it.
\label{ex:example2}
\eex

\bex
Consider the family of graphs $J_k$, for odd $k$ only, in Figure \ref{fig:examples}(c). For each vertex $a_t$ of $J_k$, 
the pair made of $a_t$ and the vertex $a_v$ following $a_t$ in the clockwise direction  satisfies 
$\mathcal{B}_2=\{b_v\}\neq\emptyset$ and $\mathcal{B}_1=\{b_t,b_{t\oplus k}\}$, where $t\oplus k$ is defined as
$t+k$ when $t\leq k$ and $t-k$ when $t>k$. Thus, with the notation
$a_p$ for the $A$-vertex following $a_t$ in the counterclockwise direction, the  
$1$-witnesses $a_p$ and $a_{p+k}$ of $a_v,a_t$ have strictly overlapping $N_1$-neighborhoods. 
By Corollary \ref{cor:notStick}$(i)$, $J_k$ is not 
a Stick graph.
\label{ex:Jk}
\eex

\br 
Note that the graphs in Figure \ref{fig:examples}(a,c) are grid intersection graphs, whereas the one in Figure \ref{fig:examples}(b) 
does not belong to GIG. Our certificate is still useful to decide the graph is not a Stick graph, since we do not have a 
simple way to show that a graph is not  a grid intersection graph.
\er

\section{Chordal Bipartite Graphs and  Stick graphs}\label{sect:holes}

{\em Chordal bipartite graphs} are the bipartite graphs with no induced long cycle (or {\em hole}), {\em i.e.} with no hole
of even size $C_{2k}$, $k\geq 3$. The relationship between chordal bipartite graphs and Stick graphs is not established, but
is interesting, due to the following two observations from the literature:

\begin{enumerate}
 \item the two maximal subclasses of Stick graphs known up to now are chordal bipartite, as shown in \cite{duran2003} for 
 the bipartite complements of circle graphs and in \cite{shrestha2010orthogonal} for 2-DOR graphs 
 (or bipartite complements of circular arc graphs).
 \item the non-Stick graphs  proposed in the literature, and more precisely in \cite{chaplick2018grid,luca2018recognition}, 
 as well as the graphs we proposed in the previous section contain at least one even hole each. 
\end{enumerate}

It is known that not all chordal bipartite graphs are Stick graphs, since chordal bipartite graphs are
not necessarily grid intersection graphs \cite{chandran2011chordal}. Reducing therefore
our analysis to bipartite chordal graphs that belong to GIG (denoted ChordalGIG), we study the place of 
ChordalGIG in the diagram in Figure \ref{fig:classes}.  

The remarks above easily imply that ${\rm Bip}\overline{\rm Circle}\subsetneq {\rm ChordalGIG}$ and 2-DORG$\subsetneq$ChordalGIG.
These are the only classes in the diagram included in ChordalGIG, as shown in the following result:

\begin{figure*}[t!]
\vspace*{-2cm}
    \centering    
    \begin{subfigure}{0.3\textwidth}
  \centering
  \includegraphics[width=5cm]{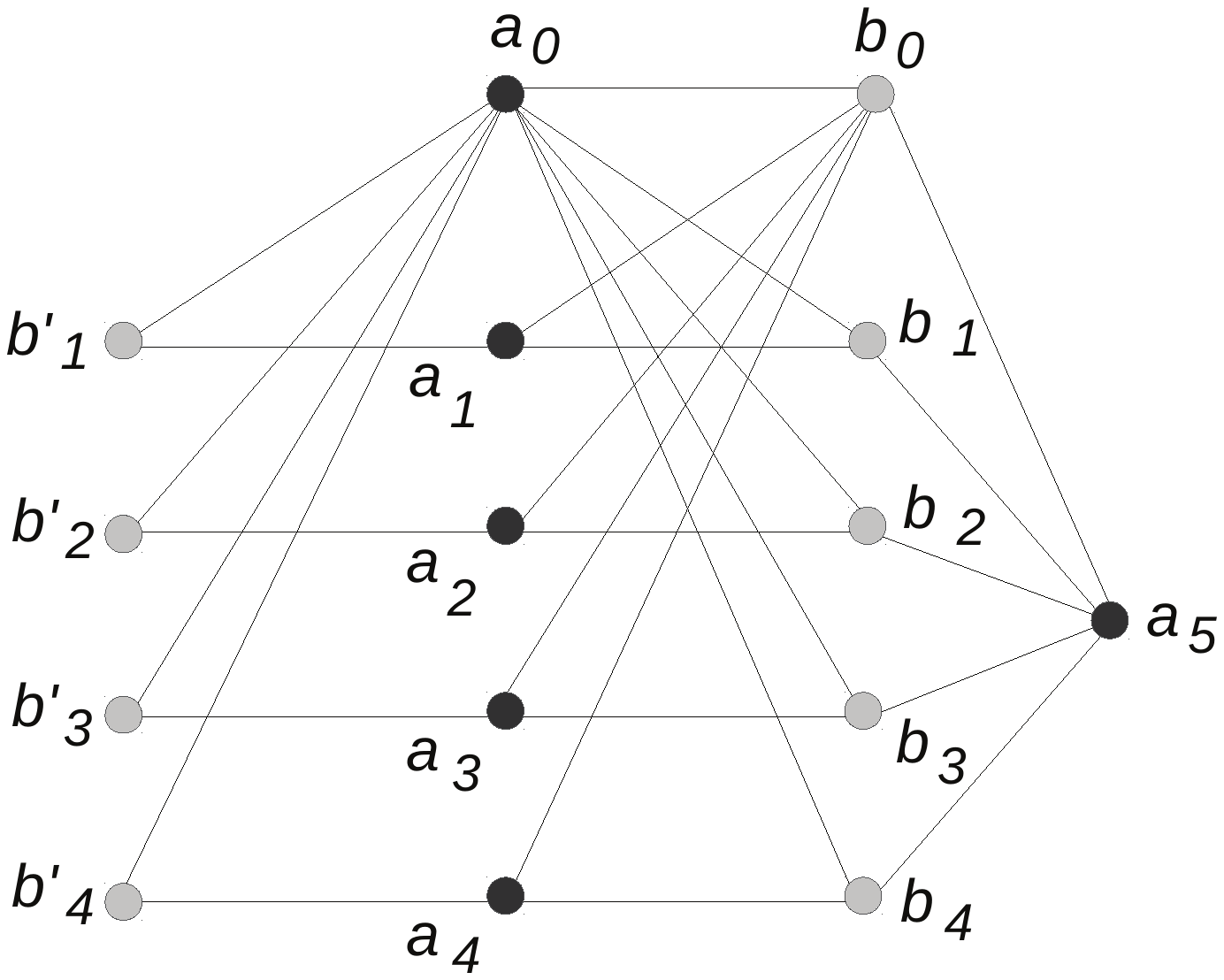}
  \label{fig:basis4}
  \caption{\small The graph $Py2$.}
  \end{subfigure}
\begin{subfigure}{0.60\textwidth}
\vspace*{0.2cm}

 \centering
  \includegraphics[width=7cm]{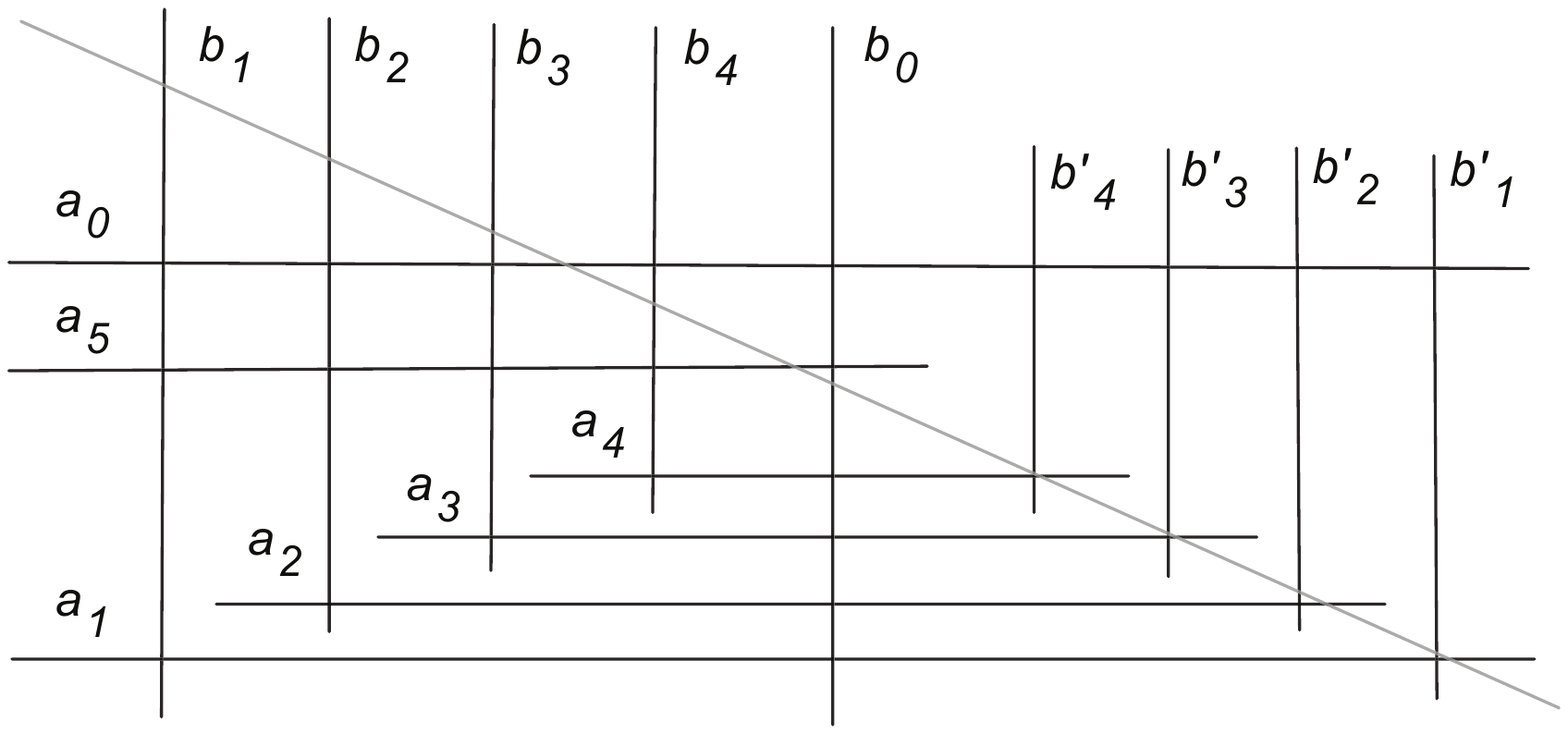}

  \label{fig:StabGIG}
  \caption{\small \vspace*{1cm} Its representation as a stabbable grid intersection graph.}
\end{subfigure}
\vspace*{-1cm} 
\caption{\small The graph $Py_2$ and one of its representations as a stabbable grid intersection graph. The notation for vertical
and horizontal segments is (exceptionally) modified in order to have up rays when the $B$-segments are transformed into rays.}
\label{fig:Py2}
\end{figure*}

\bprop
${\rm ChordalGIG}$ strictly overlaps both classes BipHook and Stick.
\label{prop:HK}
\eprop

\begin{proof}
It is clear that ${\rm Stick}\not\subset {\rm ChordalGIG}$, and thus that ${\rm BipHook}\not\subset {\rm ChordalGIG}$, since
even holes have Stick representations but are not chordal graphs.

The graph $Py2$ in Figure  \ref{fig:Py2}(a) is a chordal bipartite graph (easily verified), and is a grid intersection graph
as shown by the representation in Figure  \ref{fig:Py2}(b), where the line with slope -1 should be ignored.

We show that $Py2$ is not a BipHook graph. To this end, we investigate the possible BipHook re\-presentations of $Py2$,
and show that none of them succeeds to place a hook for each vertex so that the intersections between hooks define the edges of $Py2$.

For symmetry reasons, we may assume that 1) the centers $b_1,b_2,b_3,b_4$ of the hooks representing the vertices with the same
names are in this order on the ground line, {\em i.e.} $b_1\prec b_2\prec b_3\prec b_4$; and 2) among them, at least two are placed 
above the center of the hook $a_5$, {\em i.e.} $b_1\prec b_2\prec a_5$ (otherwise, switch the hooks above and below $a_5$,
as well as the vertical and horizontal segments of each of them). All these hooks $b_i$ intersect the hook $a_5$ on its vertical
segment (if they precede $a_5$) or on its horizontal segment (otherwise).
We first prove three affirmations. The notation $x\prec (y_1\ldots y_p)\prec u$ means $x\prec y_i\prec u $ for all $i$.
\medskip

(P1) Let $2\leq h\leq 4$. If $b_h\prec a_5$ and $a_h\prec a_5$, then $b_{h-1}\prec (a_h b'_h a_0 b_0)\prec \min\{b_{h+1},a_5\}$, 
where $\min\{b_{h+1},a_5\}=a_5$ by definition when $h=4$. 
\medskip

We have that $a_5b_j\in E$ for $1\leq j\leq 4$. Then $a_h\prec b_{h-1}$  and $a_hb_h\in E$ imply that the vertical segment of
hook $b_h$ intersects the hook $b_{h-1}$, a contradiction. Similarly,  $b_{h+1}\prec a_h$  and $a_hb_h\in E$ imply
that the vertical segment of the hook $a_h$ intersects the hook $b_{h+1}$, another contradiction. Then, since $a_h\prec a_5$ by hypothesis,   
we deduce $b_{h-1}\prec a_h\prec \min\{b_{h+1},a_5\}$. The same limits hold, for similar reasons, for the center $b'_h$, 
since the hook $b'_h$ intersects the hook $a_h$ but not the hooks $b_{h-1}, b_{h+1}, a_5$. Furthermore, the segments of 
$b'_h$ cannot cross $a_5$, $b_{h-1}$ or $b_{h+1}$, but must intersect $a_0$, which implies that $b_{h-1}\prec a_0\prec b_{h+1}$. 
Finally, $b_0$ must intersect $a_h$ which cannot cross $a_5$, $b_{h-1}$  or $b_{h+1}$, so the same limits hold for $b_0$.
\medskip

(P2) We necessarily have $b_1\prec b_2\prec b_3\prec b_4\prec a_5.$ 
\medskip

In the contrary case, we have $b_1\prec b_2 \prec a_5$ and $a_5\prec b_4$. By (P1), we cannot have $a_2\prec a_5$, since then
$b_1\prec a_0\prec \min\{b_3,a_5\}$ and the horizontal segment of the hook $a_0$ intersects the hook $a_5$ before reaching the
hook $b_4$. Thus $a_5\prec a_2$, and even $b_4\prec a_2$ since the horizontal segment of $a_5$ intersects the vertical segment of $b_4$
and $a_5\prec a_2\prec b_4$ would imply that the hooks $a_2$ and $a_5$ intersect. Similarly to the proof of (P1), we now deduce
successively that $a_4, b'_4, a_0$ are between $\max\{a_5,b_3\}$ and $a_2$. If $\max\{a_5,b_3\}\prec a_0\prec b_4$ then
the hook $a_0$ cannot intersect the hook $b_2$ (because of the horizontal segment of the hook $a_5$), and if $b_4\prec a_0$
then $a_0$ cannot intersect $b'_3$ (which is confined either between $b_2$ and $a_5$ or between $a_5$ and $b_4$). In all cases
we have a contradiction.
\medskip

(P3) One cannot have $b_i\prec b_j\prec b_k$ and $a_k\prec a_j\prec a_i$, for $1\leq i< j< k\leq  4$. 
\medskip

Note that hooks $b_t$ and $a_t$ intersect, for $t\in\{i,j,k\}$ and, by (P2), we have $b_1\prec b_2\prec b_3\prec b_4\prec a_5$.
Moreover, since $a_k\prec a_j$ and the hook $a_k$ intersects the hook $b_k$, we cannot have $a_j\prec b_k$, since in this case the 
hook  $a_j$ intersects the hook $a_k$ before the hook $b_j$, a contradiction. But then $b_k\prec a_5\prec a_j\prec a_i$, since the hook $b_k$ must 
intersect $a_5$ but not $a_j$. Note that $a_k$ may be placed anywhere between $b_j$ and $a_j$.

Assuming (P3) is false, consider the maximum triple $(i,j,k)$ in lexicographic order, such that 
$b_i\prec b_j\prec b_k$ and $a_k\prec a_j\prec a_i$. We then deduce that:
\begin{itemize}[label=$\bullet$]
\item $k=4$. Otherwise, $(i,j,k)=(1,2,3)$ and we have either $b_4\prec a_3$ which implies $a_4\prec a_3$ (since $a_4b_4\in E, a_3b_4\not\in E$) and thus $(2,3,4)$ is a configuration satisfying
the hypothesis in (P3) that should have been chosen instead of $(1,2,3)$, a contradiction; or $a_3\prec (b_4a_4) \prec a_2$
(since $a_4b_4\in E, a_2b_4\not\in E$) and thus configuration $(1,2,4)$ should have been chosen instead of $(1,2,3)$, a contradiction.
\item $b_j\prec a_0$, otherwise the hook $a_0$ cannot intersect the hook $b_4$ because of $b_3$.
\item $i=1$. 
If, by contradiction, $i\neq 1$, then $(i,j)=(2,3)$ whereas $k=4$ as proved above. Consider first the case where $a_1\prec a_5$. It implies that 
$a_1\prec b_2$ (since $a_1b_1\in E$), and thus from $a_1b_0\in E$ we deduce that either $b_0\prec b_2$ or $a_2\prec b_0$ (otherwise
the hook $b_0$ intersects the hook $b_2$ before the hook $a_1$). In both cases, the hook $b_0$ cannot intersect the hook $a_4$.
Consider now the case where $a_5\prec a_1$. Then $a_2\prec a_1$, since otherwise the hook $a_1$ intersects the hook $b_2$ before
it intersects the hook $b_1$.
Now, $b'_2, b'_3$ must avoid $a_5$ and cannot be placed above $a_5$, since then their intersection with $a_2,a_3$ respectively
would be impossible because of $a_5$. Thus $(a_5b'_4)\prec b'_3\prec b'_2\prec a_1$, and the horizontal segment
of each hook $b'_h$, $h=2, 3, 4$, 
ends before $a_{h-1}$. Then there is no way to place $a_0$ so that the hook $a_0$  intersects them all.
\item $j\neq 2,3$. By contradiction assume that $j\in\{2,3\}$ is possible. By the choice of $(i,j,k)$, one cannot have $a_5\prec a_2$ and $a_5\prec a_3$ since then we necessarily have
$a_4\prec a_3\prec a_2\prec a_1$ and we have proved that the configuration $(2,3,4)$ is not possible.
Let thus $q\in\{2,3\}$, $q\neq j$. Then $a_q\prec a_5$ and by (P1) we deduce that  $b_{q-1}\prec a_0\prec \min\{b_{q+1},a_5\}$.
But then, since $2\leq q\leq 4$, we deduce $b_1\preceq b_{q-1}\prec a_0\prec \min\{b_{4},a_5\}$, implying that $a_0$ cannot intersect 
$b'_j$, which satisfies $a_5\prec b'_j$ since $b'_j$ must intersect $a_j$ but not $a_5$.
\end{itemize}

We then deduce that (P3) holds. We now finish the proof following the steps below:

\begin{itemize}
\item at least one hook center $a_i$ satisfies $a_5\prec a_i$. In the contrary case, affirmation (P1) with $h=2$ and $h=4$ 
requires that $a_0$ be placed both before and after $b_3$, which is not possible.

\item at least two  hook centers $a_i$ satisfy $a_5\prec a_i$. In the contrary case, we may assume that the unique
$a_i$ with $a_5\prec a_i$ satisfies $i\in\{2,4\}$. Otherwise the reasoning above with $h=2$ and $h=4$ leads to a contradiction again.
Now, if $i=4$ then affirmation (P1) with $h=2$ and $h=3$ implies that $b_2\prec (a_0b_0)\prec b_3$ and thus
$b_0$ cannot intersect $a_1$ (which satisfies $a_1\prec b_2$) since it should also intersect $b_2$, a contradiction. And if
$i=2$ then a similar reasoning with $h=3$ and $h=4$ yields another contradiction.

\item exactly two  hook centers $a_i$ satisfy $a_5\prec a_i$, and one of them is $a_4$. If three or more 
hook centers are placed after $a_5$, then the only way to avoid wrong intersections is to have them in the order
forbidden by (P3), a contradiction. Thus exactly two hook centers $a_i$ and $a_j$ satisfy $a_5\prec a_i\prec  a_j$. 
If $a_4$ is not among them, then necessarily $a_4\prec a_5\prec a_i\prec a_j$ whereas we must have $b_j\prec b_i\prec b_4$
in order to avoid wrong intersections, and thus (P3) is contradicted.
\end{itemize}

But the previous configuration is not realizable either. Let $i\neq 4$ such that $a_5\prec a_4\prec a_i$. Then $a_5\prec b'_4\prec a_i$,
since the hook $b'_4$ intersects $a_4$ but not $a_5$. Let $j\in\{2,3\}\setminus\{i\}$. By (P1), $b_1\prec b'_j\prec a_5$.
Thus there is no way to place $a_0$ such that it intersects both the hooks $b'_4$ and $b'_j$ but not the hook $a_5$. 
\end{proof}

As shown below, the graph $Py2$ sharply cuts between BipHook and all its known superclasses, since it belongs
to all of them. Trying to go even closer to Stick graphs, we do not know 
whether there exist graphs that belong both to BipHook and ChordalGIG, but that are not Stick graphs. 
\bigskip

\bprop
The classes {\rm 3-dim GIG} $\cap {\rm StabGIG} \cap {\rm SegRay} \cap {\rm ChordalGIG}$ and {\rm BipHook} strictly overlap.
\label{prop:ToutMaisPasBipHook}
\eprop

\begin{proof}
The Stick graph $C_6$, which also belongs to BipHook, shows that ${\rm BipHook}\not\subset$3-dim GIG$\cap {\rm StabGIG} \cap {\rm SegRay} \cap {\rm ChordalGIG}$.
We show that $Py2$ belongs 3-dim GIG$\cap {\rm StabGIG} \cap {\rm SegRay}$, and thus the
conclusion follows from Proposition \ref{prop:HK}.

Figure \ref{fig:Py2}(b) gives a GIG representation for $Py2$ with a line crossing all the segments, thus showing
that $Py2$ is a stabbable GIG graph. All the $B$-segments  may be transformed into 
up rays without modifying the intersections, thus showing that $Py2$ belongs to SegRay. 
 
 Furthermore, consider the relation $\leq_{Py2}$ defined on the set of vertices $A\cup B$ of $Py2$, such that 
 $a\leq_{Py2} b$ for each $a\in A$ and $b\in B$ such that $ab\in E$. The three following linear orders form a realizer of $\leq_{Py2}$:
 \medskip
%
%
%

 $a_5\leq_1 a_4\leq_1 a_0\leq_1 b_4\leq_1 b'_4\leq_1 a_3\leq_1 b_3\leq_1 b'_3\leq _1 a_2\leq_1 b_2\leq_1 a_1\leq_1 b_1\leq_1 b_0\leq_1 b'_2\leq_1 b'_1$
 
 $a_5\leq_2 a_1\leq_2 a_0\leq_2 b_1\leq_2 b'_1\leq_2 a_2\leq_2 b_2\leq_2 b'_2\leq_2 a_3\leq_2 b_3\leq_2 a_4\leq_2 b_4\leq_2 b_0\leq_2 b'_3\leq_2 b'_4$
 
 $a_3\leq_3 a_2\leq_3 a_0\leq_3 b'_3\leq_3 a_4\leq_3 b'_2\leq_3 a_1\leq_3 b'_4\leq_3 b'_1\leq_3 a_5\leq_3 b_0\leq_3 b_4\leq_3 b_3\leq_3 b_2\leq_3 b_1$
 \medskip
 
 Thus $Py2$ is a 3-dimensional GIG.
\end{proof}

We end the study of inclusions and overlaps between $ChordalGIG$ and the other classes with this result.

\bprop
{\rm ChordalGIG} strictly overlaps each of the  classes {\em SegRay} and {\em StabGIG}.
\label{prop:SegStab}
\eprop

\begin{figure}[t]
\vspace*{-3cm}
 \centering
 \includegraphics[width=18cm]{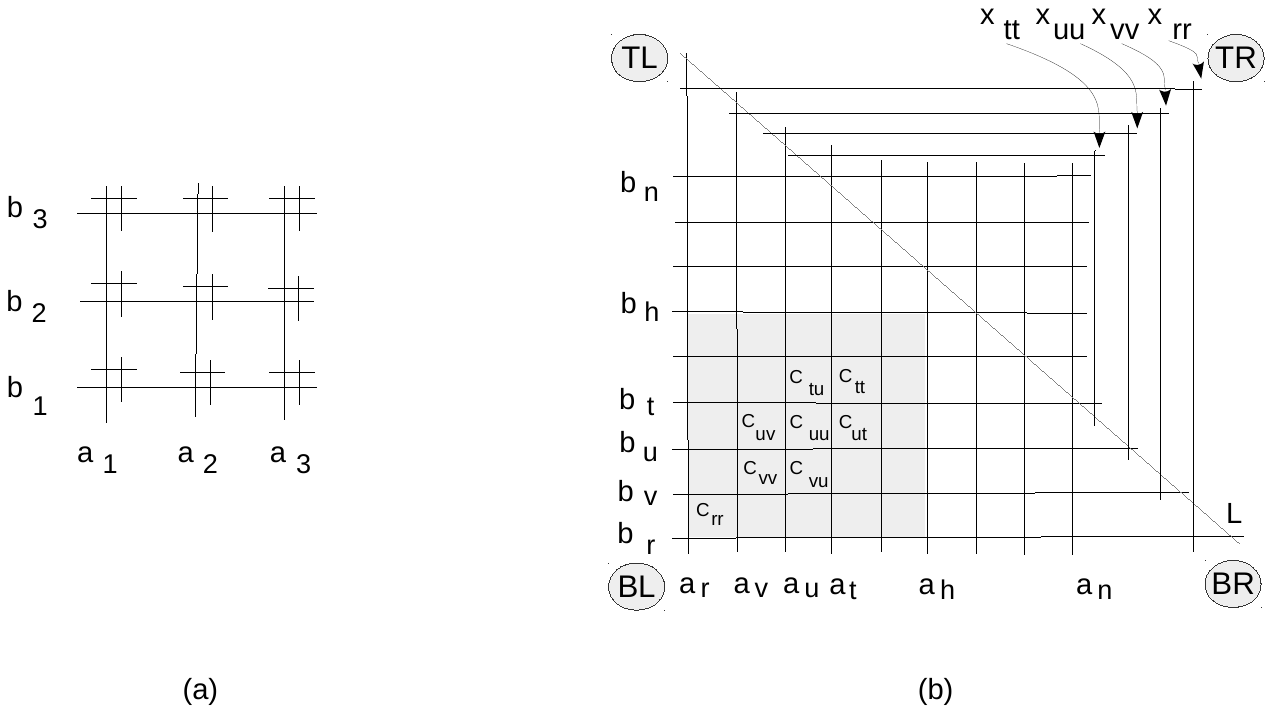}
\vspace*{-2cm}

 \caption{\small  \label{fig:otachi} (a) Grid representation of the graph $O_{3,3}$, issued from \cite{otachi2007relationships}. 
 (b) Unsuccessful attempt to define a stabbable grid representation of $O_{n,n}$ (see Proposition \ref{prop:SegStab}). Only the
 significant rows and columns are drawn.}
\end{figure}

\begin{proof}
 The graph $O_{3,3}$, proposed in \cite{otachi2007relationships}, whose grid representation is given in Figure \ref{fig:otachi}(a) is a chordal
 graph, but not a segment-ray graph. To see this, notice that the symmetry allows to assume without loss of generality
 that the segments $a_i$ and $b_i$, for $1\leq i\leq 3$, intersect as in the figure. Then $a_1, a_2, a_3$ may be immediately transformed into 
 up rays. 
 
 Use the notation $a_{ij}$ and $b_{ij}$ to designate the two short vertical and respectively horizontal segments, 
 and their associated vertices, drawn at the intersection of the segments $b_i$ and $a_j$ in Figure \ref{fig:otachi}(a). Then
 it is possible to move the segments $a_{j1}$, $j=1,2$, towards left while lengthening $b_{j1}$, and similarly for the
 segments  $a_{j3}$ and $b_{j3}$, $j=1,2$, but towards right, in order to reach a position allowing the transformation of
 $a_{j1}$ and $a_{j3}$ $(j=1, 2, 3)$ into up rays. Furthermore, by moving $b_{12}$ lower that $b_{1}$, the segment $a_{12}$ may be
 moved to the extreme right (or, equivalently, to the extreme left) so that it may be transformed into an up ray. But there is no way
 to move $a_{22}$ and $b_{22}$ in order to transform $a_{22}$ into an up ray. Thus the graph $O_{3,3}$ does not belong to SegRay.
 
 Whereas this graph is stabbable, larger graphs $O_{n,n}$ built on the same principles are not stabbable. Let $n\geq 15$ and 
 assume, due to symmetry, that the segments $a_i, b_i$ ($1\leq i\leq n$) are placed in this order from left to right (for $a_i$)
 and from bottom to top (for $b_i$) in the stabbable grid representation we are looking for.
 Think first of $O_{n,n}$ as a matrix with $n-1$ rows and $n-1$ columns defining cells. See Figure \ref{fig:otachi}(b).  Let $C_{ij}$ be the cell whose bottom left 
 corner is the intersection of $b_i$ and $a_j$.
 Note that the straight line $L$ used for stabbing the segments is neither horizontal nor vertical, and necessarily avoids at least a 
 quarter of the matrix, {\em i.e.} a region  $R$ of $h\times h$ cells, with $h=\lfloor \frac{n-1}{2}\rfloor$, one of whose corners coincide with a corner of the matrix.
 Without loss of generality, we assume $R$ is made of the cells $C_{ij}$ with $1\leq i,j\leq h$ of $O_{n,n}$,
 and $L$ intersects $a_1$ ($b_1$) above the cell $C_{h1}$ (to the right of the cell $C_{1h}$). Then, by the definition
 of $R$, the line $L$ intersects none of the cells in $R$, but may contain only the top right corner of $R$. 
 See Figure \ref{fig:otachi}(b). 
 
 Then, recalling that $a_{ij}$ and $b_{ij}$ must intersect $L$ for all $i$ and $j$, we deduce that:
 
 \begin{itemize}
  \item For each $ij$ such that $1\leq i,j\leq h$, $a_{ij}$ (respectively $b_{ij}$) is placed to the left of $a_1$ or to 
  the right of $a_n$ (respectively above $b_n$ or below $b_1$). Thus the intersection  point $x_{ij}$
  of $a_{ij}$ and $b_{ij}$ is placed in one of the four regions: bottom-left (BL), bottom-right (BR), top-left (TL) and 
  top-right (TR), each defined as the intersection of two of the half-planes defined above by some $a_k$, $k\in\{1,n\}$ and
  some $b_l$, $l\in\{1,n\}$.
  \item None of the regions BL, RL, TL accepts more than one intersection point $x_{ii}$ such that $1\leq i\leq h$. 
  Indeed, in the contrary case, with $x_{ii}$ and $x_{jj}$ in the same region ($i<j$), once the segments 
  $a_{jj}$ and  $b_{jj}$ are placed such as they intersect $L$, each placement of $a_{ii}$ and $b_{ii}$ ensuring that they
  intersect $L$ implies a wrong intersection between $a_{ii}$ and $b_{jj}$, or between $a_{jj}$ and $b_{ii}$.
  \item Region TR accepts as many $x_{ii}$ as needed (assuming the line $L$ is close enough to the top right corner of $O_{n,n}$;
  otherwise all the $x_{ii}$ cannot be placed and we are already done). Moreover, the segments $b_{ii}$ are ordered in increasing order of their 
  $i$ from top to bottom, whereas the segments $a_{ii}$ are ordered in increasing order of their $i$ from right to left.
  \end{itemize}
  
  Since $n\geq 15$, we have $h\geq 7$, and the deductions above imply that at least four of the points $x_{ii}$ are
  placed in the region TR. Let $r<v<u<t$ be four indices such that $x_{rr}, x_{vv}, v_{uu}$ and $x_{tt}$ belong to TR. Now,
  $x_{tu}$  cannot belong to TL or BR. Indeed,  if $x_{tu}$ belonged to TL, then the segment $a_{tu}$  
  would be situated to the left of $a_r$, whereas $b_{tu}$ would be situated to the right of $a_r$ since it intersects $a_u$ 
  below  $b_{rr}$. The intersection between $a_{tu}$ and $b_{tu}$ is impossible in TL. The reasoning is similar for
  BR. Thus $x_{tu}$ cannot belong to TL or BR, and a similar reasoning holds for $x_{ut}, x_{vu}, x_{uv}$.
  
  But now it is easy to check that BL does not accept more than one intersection point among $x_{tu}$, $x_{ut}$, $x_{uv}$ and 
  $x_{vu}$. As an example, if $x_{uv}$ belongs to BL, and we try to place $x_{vu}$ in BL too assuming $b_{uv}$ and
  $a_{uv}$ are already placed, then $b_{vu}$ must intersect  
  $L$ (to the right of $a_v$), $a_{vu}$ (to the left of $a_v$),  $a_{u}$ (between $b_1$ and $b_{uv}$) but not $a_v$. 
  This is not possible, and the same type of reasoning applies for the other pairs of intersection points. We then deduce
  that at least three of the four points belong to TR, two of which are either $x_{uv}$ and $x_{vu}$, or $x_{tu}$ and $x_{ut}$.
  But now, once one of the two points is placed in TR, there is no way to place the second one so as to obtain exactly
  the required intersections.

  Thus $O_{n,n}$ is not stabbable for $n\geq 15$. As $O_{n,n}$ is a chordal grid intersection graph, it shows that
  ChordalGIG$\not\subset$StabGIG. And, as before, long holes show that none of the classes SegRay and StabGIG is 
  included in ChordalGIG.
\end{proof}

\br Since it is a part of GIG, the class ChordalGIG contains graphs of dimension at most 4. However, although we are convinced
that the graphs $O_{n,n}$, for $n\geq 3$, are of dimension equal to 4, we were not able to prove it. This explains the
dotted separation between ChordalGIG and the 3-dimensional region in the diagram in Figure \ref{fig:classes}, marked with a
''?''. Note that if all the graphs $O_{n,n}$ were of dimension 3 (they are certainly of dimension at least 3), then
$O_{15,15}$ would solve the other interrogation point in Figure \ref{fig:classes}, since  $O_{15,15}$ would be a 
graph from 3-dimGIG but not from  StabGIG. In conclusion, $O_{15,15}$ solves one of the questions left open in the
diagram, probably the one to the right. 
\label{rem:Chordal3dim}
\er

\section{Conclusion}\label{sect:conclusion}

The classes of grid intersection graphs we studied in this paper are challenging classes of graphs, 
still very little known. We focused here on Stick graphs, with the aim of perceiving the frontiers of this class: we proposed
a new maximal subclass, new forbidden subgraphs together with certificates to decide that a graph is not a Stick graph,
and we showed that the class Stick strictly overlaps the intersection of all its known superclasses, even when it 
is reduced to graphs without holes. 

The main open question about Stick graphs, but also about each of its superclasses in Figure \ref{fig:classes} except GIG, concerns
the complexity of recognizing the class. Two less important, but interesting, open questions are directly raised by the study
in this paper: find a 4-dimensional chordal grid intersection graph (or show that some $O_{n,n}$ is of dimension equal to 4) 
and thus solve the question we added in the class diagram; and
find a chordal grid intersection graph which is BipHook but not Stick, thus providing an example of a non-Stick graph even closer 
to Stick than $Py2$.

\bibliographystyle{plain}
\bibliography{Stick}
\end{document}